\documentclass[fleqn,usenatbib]{mnras}

\usepackage{newtxtext,newtxmath}

\usepackage[T1]{fontenc}
\usepackage{ae,aecompl}

\usepackage{graphicx}	
\usepackage{amsmath}	
\usepackage{amssymb}	
\usepackage{color}

\newcommand{\lya}{Lyman-$\alpha$ }

\title[Reionization effect on the \lya forest]{Impact of inhomogeneous reionization on the \lya forest}

\author[P. Montero-Camacho et al.]{
Paulo Montero-Camacho,$^{1,2}$\thanks{E-mail: monterocamacho.1@osu.edu (PMC)}
Christopher M. Hirata,$^{1,2,3}$
Paul Martini$^{1,3}$
and  
\newauthor{Klaus Honscheid$^{1,2}$}
\\
$^{1}$Center for Cosmology and Astroparticle Physics, The Ohio State University, 191 West Woodruff Lane, Columbus, Ohio 43210, USA. \\
$^{2}$Department of Physics, The Ohio State University, 191 West Woodruff Lane, Columbus, OH 43210, USA\\
$^{3}$Department of Astronomy, The Ohio State University, 140 West 18th Avenue, Columbus, OH 43210, USA}

\date{Accepted XXX. Received YYY; in original form ZZZ}

\pubyear{2017}

\begin{document}
\label{firstpage}
\pagerange{\pageref{firstpage}--\pageref{lastpage}}
\maketitle

\begin{abstract}

The \lya forest at high redshifts is a powerful probe of reionization. Modeling and observing this imprint comes with significant technical challenges: inhomogeneous reionization must be taken into account while simultaneously being able to resolve the web-like small-scale structure prior to reionization. In this work we quantify the impact of inhomogeneous reionization on the \lya forest at lower redshifts ($2 < z < 4$), where upcoming surveys such as DESI will enable precision measurements of the flux power spectrum. We use both small box simulations capable of handling the small-scale structure of the \lya forest and semi-numerical large box simulations capable of representing the effects of inhomogeneous reionization. We find that inhomogeneous reionization could produce a measurable effect on the \lya forest power spectrum. The deviation in the 3D power spectrum at $z_{\rm obs} = 4$ and $k = 0.14 \ \textup{Mpc}^{-1}$ ranges from 19 -- 36$\%$, with a larger effect for later reionization. The corrections decrease to 2.0 -- 4.1$\%$ by $z_{\rm obs} = 2$. The impact on the 1D power spectrum is smaller, and ranges from 3.3 -- 6.5$\%$ at $z_{\rm obs}=4$ to 0.35 -- 0.75$\%$ at $z_{\rm obs}=2$, values which are comparable to the statistical uncertainties in current and upcoming surveys. Furthermore, we study how can this systematic be constrained with the help of the quadrupole of the 21 cm power spectrum.

\end{abstract}




\section{Introduction}

In the standard model of cosmology, the diffuse gas in the Universe has undergone a complex thermal history marked by several important transitions. At early times, the gas was hot, dense, ionized, and fully coupled to the thermal radiation in the early Universe. As cosmic recombination occurred at $z\sim 1100$, the gas became transparent; perturbations from this epoch are directly observable to us as fluctuations in the cosmic microwave background (CMB) temperature and polarization. During the subsequent epoch -- the cosmic Dark Ages -- perturbations in the gas and dark matter grew by gravitational instability, which eventually went non-linear and began to form collapsed objects. Some of these objects emitted ionizing radiation, triggering cosmic reionization. This transition, currently believed to have occurred at $z_{\rm re} = 7.7\pm 0.8$ \citep{2018arXiv180706209P}, also must have re-heated the intergalactic medium (IGM) 
%
to temperatures up to a few$\,\times 10^4 \ \textup{K}$ after the passing of an ionization front \citep{2016ARA&A..54..313M, 2018arXiv180709282D}. After reionization it is possible to probe the IGM using the absorption of neutral hydrogen in quasar spectra from overdense regions -- the ``\lya forest''. Each quasar spectrum probes a 1-dimensional skewer through the Universe, with the fraction of flux transmitted at each observed wavelength being inversely related to the gas density at the corresponding redshift\footnote{More precisely, the optical depth is proportional to the neutral hydrogen density, which in turn depends on the density and temperature of the gas and the photoionization rate. This is then smeared by the Doppler width of the Lyman-$\alpha$ line.}.

In addition to probing the astrophysics of the IGM, the \lya forest has become an important tool for observational cosmology. It probes a redshift range ($2<z<5$) where conventional large-area galaxy surveys are very difficult due to the enormous luminosity distance and the redshifting of bright nebular emission lines into the infrared. In contrast, while bright background quasars are rare, each one provides information from the many structures along their line of sight. The \lya\ forest is also in some ways simpler to model than galaxies: most of the forest consists of moderately overdense gas, where the physics is dominated by gravitational instability rather than feedback processes \citep{1994ApJ...437L...9C, 1996ApJ...457L..51H, 1997ApJ...486..599H, 2000ApJ...532..118M, 2015MNRAS.446.3697L}; see \citet{2016ARA&A..54..313M} for a recent review. The \lya forest has been used to study reionization of hydrogen and helium \citep{2002AJ....123.1247F, 2009ApJ...706L.164C, 2009ApJ...694..842M, 2010MNRAS.408...57P, 2011MNRAS.410.1096B, 2013MNRAS.435.3169C, 2015MNRAS.446..566M,  2015MNRAS.447.2503G, 2015MNRAS.447..499M, 2015MNRAS.452..261C, 2015ApJ...811..140B, 2016MNRAS.463.2335N,
2017ApJ...847...63O, 2018arXiv181011683O, 2018arXiv180804367W}; the matter power spectrum (particularly on small scales that are too non-linear in galaxy surveys; \citealt{2008PhRvL.100d1304V, 1998astro.ph.10142W, 2003AIPC..666..157W}); constrain cosmological parameters and the late evolution of the universe \citep{2002APh....17..375H, 2005PhRvD..71j3515S, 2006JCAP...10..014S, 2006MNRAS.369.1090C, 2018arXiv181203554C}; and measure the neutrino mass via its effect on the growth of structure 
\citep{2015JCAP...11..011P, 2017JCAP...06..047Y}.

Motivated by both cosmology and astrophysics, there has been a growing number of observed \lya\ forest sightlines in the range $2 < z <5$ \citep{1994ApJS...91....1B, 1998ApJ...495...44C,
1999ApJ...520....1C,  2002ApJ...581...20C,
1998ARA&A..36..267R,
2000ApJ...543....1M, 2006ApJS..163...80M, 2010ApJ...718..199L, 
2011JCAP...09..001S, 2013JCAP...04..026S,
2011AJ....142...72E, 2013AJ....145...10D, 2014JCAP...05..027F, 2015A&A...574A..59D,
2017A&A...608A.130D, 2017A&A...603A..12B,  
2017MNRAS.466.4332I,
2018ApJ...852...22W, 
2018ApJS..237...31L}.
A range of statistical measures of the \lya forest have been used; for cosmology purposes, the most common has been the 1D correlation function or power spectrum (that is, the power spectrum of individual skewers, treated as 1D random fields). With the Baryon Oscillation Spectroscopic Survey (BOSS), the number of sightlines became great enough to extract cosmological parameters from the 3D correlation function of the \lya forest (that is, using correlations between different lines of sight). BOSS also measured cross-correlations between Damped \lya systems (DLA) and the \lya forest \citep{2012JCAP...11..059F}, and measured Baryon Acoustic Oscillations (BAO) in the \lya forest \citep{2013JCAP...04..026S}. The number of observed \lya forest will soon be dramatically enhanced with the Dark Energy Spectroscopic Instrument \citep[DESI;][]{2016arXiv161100036D}.

The current paradigm for reionization is that it is an extended and inhomogeneous process, with ``bubbles'' of ionized gas forming in regions with more sources of ionizing photons, and the end of reionization is when these bubbles overlap and approach a filling factor of unity. Inhomogeneous reionization leaves its imprint in the \lya forest even after reionzation ends, because the thermal state of the gas depends on when it was reionized \citep{1994MNRAS.266..343M} and because the thermal history itself affects the distribution of the gas at the Jeans scale. Because the attractor evolution of the IGM temperature causes the memory of reionization to fade with time \citep[see][for a detailed explanation]{2016MNRAS.456...47M}, as well as the complication of helium reionization turning on at later redshifts ($z \sim 3.5$), most studies on the effect of reionization in the \lya forest have been at the highest redshifts that have a sufficient number of sightlines \citep{2003ApJ...596....9H, 2008ApJ...689L..81T, 2009ApJ...701...94F, 2014ApJ...788..175L, 2015ApJ...813L..38D, 2018MNRAS.477.5501K}.

The goal of this paper is to compute a first estimate of the effect of inhomogeneous reionization in the \lya forest at the {\em lower} redshifts ($2<z<4$) relevant to precision cosmology programs such as DESI. The remnants of hydrogen reionization are expected to be small, but in the DESI era even small effects in the \lya forest power spectrum will be important. Due to the enormous dynamic range in spatial scales, we use two types of simulations. Our approach here is analogous to the recent work by \cite{2018arXiv181011683O}, where the effect of inhomogeneous reionization in the 1D \lya forest power spectrum was computed at higher redshifts ($4 < z < 6$). Our small-scale simulations use full hydrodynamics to follow the dynamics of gas down to below the Jeans scale, but with an ionization history set by hand. These use the same modification of {\sc Gadget-2} \citep{2005MNRAS.364.1105S} that \citet{2018MNRAS.474.2173H} used to study the streaming velocity effect in the \lya\ forest. The large-scale boxes use a semi-numerical approach ({\sc 21cmFAST}; \citealt{2007ApJ...669..663M, 2011MNRAS.411..955M}) to model the ionized bubbles and their correlation with large-scale structure.

Besides estimating the change $\Delta P_{{\rm Ly}\alpha}(k,z)$ in \lya forest power spectrum due to reionization effects, we consider how $21$ cm observations could be used to predict $\Delta P(k,z)$ and mitigate this systematic effect. The H{\sc\,i} 21 cm hyperfine transition is a potentially powerful probe of reionization, and global measurements of the signal have already ruled out some models \citep{2017ApJ...847...64M, 2017ApJ...845L..12S, 2018ApJ...858...54S}. We find that there is a quantitative relation between $\Delta P_{{\rm Ly}\alpha}(k,z)$ and the full history of the cross-correlation of the matter and the ionization fraction. The latter is related to the redshift-space distortion in the 21 cm power spectrum \citep{2005ApJ...624L..65B, 2012MNRAS.422..926M}. We show that the simplest approach -- using linear pertubation theory to map the $\ell=2$ quadrupole of the 21 cm signal into a matter-ionization fraction cross-power spectrum and using this to predict $\Delta P_{{\rm Ly}\alpha}(k,z)$ -- is not accurate. We thus recommend that future work consider mitigation using models for the 21 cm redshift space distortion that go beyond linear theory \citep[e.g.,][]{2012MNRAS.422..926M}.

We expect the effect studied here to be of importance for current and near-future \lya experiments. In particular, DESI will measure the three dimensional (3D) and one dimensional (1D) power spectrum of the \lya forest \citep{2016arXiv161100036D}, both of which will be altered by inhomogeneous reionization. The 1D power spectrum contains the correlation of the spatial structure of the neutral hydrogen regions along the line of sight of the same quasar \citep{2005ApJ...635..761M,2006ApJS..163...80M,2013A&A...559A..85P}, i.e. the individual skewers. In contrast, the 3D spectrum has correlations between different lines of sight \citep{2015JCAP...11..034B, 2017A&A...603A..12B, 2018JCAP...01..003F}. The fractional effect $\Delta P(k,z)/P(k,z)$ of inhomogeneous reionization is larger on large scales (smaller $k$) than at smaller scales, and is larger in the 3D \lya forest power spectrum than in 1D (because large $k$ in 3D can map into small $k$ in 1D but not the other way around). The 1D power spectrum is affected by non-linearities, even for small $k_{\parallel}$, due to small-scale processes that govern the evolution of the IGM \citep{2013A&A...559A..85P,2015JCAP...12..017A}. 

This paper is organized as follows. We introduce the relevant formalism in \S\ref{sec:formalism}. We describe the two types of simulations we use in \S\ref{sec:setup}. Then we proceed to assess the possible contamination of the \lya flux in \S\ref{sec:contamination}, and confirm that for this systematic the 3D \lya power spectrum is more affected than the 1D. In \S\ref{sec:21cmcross} we explore the use of 21 cm observables to address this contamination, and find that non-linear effects must be taken into account to appropriately utilize the link between the imprint of inhomogeneous reionization in the \lya forest and the $21$ cm quadrupole. We summarize our results and discuss directions for future work in \S\ref{sec:summary}.

\section{Conventions and formalism}
\label{sec:formalism}

Throughout this paper we use the $\Lambda$CDM cosmological parameters from \textit{Planck} 2015 `$TT + TE + EE + \textup{lowP} + \textup{lensing} + \textup{ext}$' \citep{2016A&A...594A..13P}, namely: $\Omega_{\rm b}h^2 = 0.02230$, $\Omega_{\rm m}h^2 = 0.14170$, $H_0 = 67.74 \ \textup{km} \, \textup{s}^{-1} \, \textup{Mpc}^{-1}$, $\sigma_8 = 0.8159$ and $n_s = 0.9667$. 

For convenience throughout this paper we use the dimensionless power spectrum when we refer to a given power spectrum, i.e. no $\textup{Mpc}^3$ unless explicitly stated. However, in the case of the 21 cm power spectrum the $k^3 P(k)/2\pi^2$ have units of $\textup{mK}^2$. Furthermore, we define all fluctuations as $\delta_p = p/\bar{p} - 1$, with the only exception of the neutral hydrogen fraction fluctuations where $\delta_{x_{\rm HI}} = x_{\rm HI} - \bar{x}_{\rm HI}$.

The fluctuations of the \lya forest transmitted flux are described by $\delta_{\rm F} = (F - \bar{F})/\bar{F}$, where $\bar{F}$ is the mean observed flux and the transmitted flux is normalized such that $0 \leq F \leq 1$. Then at linear order\footnote{See \cite{2015JCAP...12..017A} for a description of the linear biasing coefficients and their range of validity.} the flux fluctuations should be proportional to the matter fluctuations. Furthermore, taking into account redshift space distortions due to velocity gradients we have $\delta_{\rm F} = b_{\rm F} (1 + \beta_{\rm F} \mu^2) \, \delta_{\rm m}$, where $\delta_{\rm m}$ is the matter density contrast, $b_{\rm F}$ is the usual \lya flux bias, $\beta_{\rm F}$ is the redshift distortion parameter and $\mu = \cos\theta = k_{\parallel}/k$ the angle to the line of sight. Moreover, the corresponding power spectrum of $\delta_{\rm F}$ is denoted by $P_{\rm F} (k, \mu)$. 

In this work we explore the effect of inhomogeneous reionization in the \lya transmission,
\begin{eqnarray}
\label{eq:fluctflux}
\delta_{\rm F}  =  b_{\rm F}\, (1 + \beta_{\rm F} \mu^2) \, \delta_{\rm m} + b_{\rm \Gamma} \, \psi (z_{\rm re})\, \textup{,}
\end{eqnarray}
where $b_{\rm \Gamma}$ is the radiation bias defined by $b_{\rm \Gamma} = \partial \ln \bar{F}/\partial \ln \tau_1$ \citep{2015JCAP...12..017A,2018MNRAS.474.2173H}. Here $\tau_1$ is the optical depth that must be assigned to a patch of gas with mean density $\Delta_b\equiv 1+\delta_b= 1$ and temperature $T=10^4 \ \textup{K}$ in order for the mean transmitted flux $\bar F$ to match observations.\footnote{This is the same usage as in \citet{2018MNRAS.474.2173H}.} This bias parameter
is needed here due to the way the post-processing of our small-scale simulations works (see \S\ref{sec:setup}). As is common in Lyman-$\alpha$ forest studies, we vary the normalization of $\tau_1$ (or equivalently, we vary the ionizing background) until we obtain the correct mean transmitted flux $\bar F$. Thus, when varying any aspect of the simulation -- here the redshift of reionization -- the quantity we measure in the simulation is $\Delta\ln\tau_1$, and the corresponding change in flux $\delta_F$ has an additional factor of $b_\Gamma$. In addition, we define
\begin{eqnarray}
\label{eq:psire}
\psi(z_{\rm re})  =  \Delta  \ln \tau_1 (z_{\rm re}, \bar{z}_{\rm re}) = \ln \left[\frac{\tau_1 (z_{\rm re})}{\tau_1 (\bar{z}_{\rm re})}\right] \, \textup{,}
\end{eqnarray}
which parametrizes the variations in the transparency for an opacity cube that suddenly reionizes at $z_{\rm re}$ relative to an opacity cube that reionizes at $\bar{z}_{\rm re}$. The sign convention is such that $\psi(z_{\rm re})>0$ if gas reionized at $z_{\rm re}$ is more transparent (has higher $F$) than gas reionized at $\bar z_{\rm re}$.

The corresponding 3D \lya flux power spectrum is given by
\begin{eqnarray}
P^{\rm 3D}_{\rm F} (k, \mu, z_{\rm obs}) &\approx & b^2_{\rm F} \, (1 + \beta_{\rm F} \mu^2)^2 P_{\rm m}(k, z_{\rm obs}) \nonumber \\
\label{eq:pof}
&&+ \, 2 b_{\rm F} \, b_{\rm \Gamma} \,(1 + \beta_{\rm F} \mu^2) P_{{\rm m},\psi}(k, z_{\rm obs}) \, \textup{,}
\end{eqnarray}
where $P_{\rm m}$ is the matter power spectrum and $P_{{\rm m}, \psi}$ is the cross-power spectrum of matter and $\psi$. Note that we have neglected the third term in equation (\ref{eq:pof}), the \textit{auto}-power spectrum of $\psi$, because it is second order in $\psi$. 

In principle the problem is now reduced to devising a way to model $P_{{\rm m}, \psi}(k,z)$ and, since it involves physics from different scales, we will \textit{divide and conquer}. We begin by re-writing the cross-power spectrum $P_{{\rm m}, \psi}(k,z)$ in a form that is straightforward and numerically stable to compute from {\sc 21cmFAST} boxes -- in particular, that can be computed from a sequence of redshift outputs during reionization. We see that
\begin{eqnarray}
    && \!\!\!\!\!\!\!\!\!\!\!\!\!\!\!\!
    (2\pi)^3 \delta^{(3)}(\boldsymbol{k} -\boldsymbol{k}') P_{\rm{m},\psi} (z_{\rm obs},k)
    \nonumber \\
    & = &\!\! \int_{\mathbb{R}^3} d^3\boldsymbol{r}' {\rm e}^{-{\rm i}\boldsymbol{k}' \cdot \boldsymbol{r}'} \left\langle \tilde{\delta}_{\rm m}^* (z_{\rm obs}, \boldsymbol{k}) {\psi} (z_{\rm re}(\boldsymbol{r}'),z_{\rm obs}) \right\rangle \nonumber \\
    & = &\!\! -\int_{\mathbb{R}^3} d^3\boldsymbol{r}' {\rm e}^{-{\rm i}\boldsymbol{k}' \cdot \boldsymbol{r}'} \int_{z_{\rm max}}^{z_{\rm min}} \Bigl\langle \tilde{\delta}_{\rm m}^* (z_{\rm obs}, \boldsymbol{k}) \frac{\partial \psi(z',z_{\rm obs})}{\partial z'}
    \nonumber \\ &&~~~~ \times\, \Theta (z' -z_{\rm re}(\boldsymbol r'))  \Bigr\rangle dz' \nonumber  \\
    & =& \!\! - \int_{z_{\rm min}}^{z_{\rm max}} dz' \frac{\partial \psi}{\partial z'}(z',z_{\rm obs}) \langle
    \tilde{\delta}_{\rm m}^* (z', \boldsymbol{k})
    \tilde{x}_{\rm HI}(z', \boldsymbol{k}')   \rangle \frac{D(z_{\rm obs)}}{D(z')}. ~~~~~
    \end{eqnarray}
Here we write $\tilde{\psi}$ as an explicit Fourier transform of the real space in the second line. We then introduce the step function and apply the fundamental theorem of calculus to write $\psi(z_{\rm re}(\boldsymbol r',z_{\rm obs})$ as negative an integral of its derivative from $z_{\rm re}(\boldsymbol r')$ to $z_{\rm max}$. (Here $z_{\rm min}$ and $z_{\rm max}$ span the range of redshifts for reionization. The $\psi(z_{\rm max},z_{\rm obs})$ term has no spatial dependence and drops out for $k\neq 0$.) In the last equality, we have used the ratio of growth functions to extrapolate $\tilde\delta_{\rm m}$ from $z'$ to $z_{\rm obs}$. Also we have used the fact that $x_{\rm HI} (\boldsymbol{r}',z') = \Theta(z'-z(\boldsymbol{r}'))$, and then wrote the neutral hydrogen fraction field in Fourier space. Thus:
    \begin{eqnarray}
 P_{{\rm m},\psi} (z_{\rm obs}, k) = - \int_{z_{\rm min}}^{z_{\rm max}} \frac{\partial \psi}{\partial z} P_{\rm m, x_{HI}}(z, k) \frac{D(z_{\rm obs})}{D(z)} dz \, \textup{,}
    \label{eq:crossps}
\end{eqnarray}
We integrate from $z_{\rm max} =34.7$ to $z_{\rm min} =5.90$, such that in our default model we cover from the almost-fully-neutral stage through the end of reionization. Hence we have computed the cross-power spectrum of $\psi$ and matter by using the cross-power spectrum of matter and neutral fraction to describe how matter and bubble spatial structure are correlated at a given redshift, i.e. how patchy reionization enters the fray. Furthermore, Eq.~(\ref{eq:crossps}) directly depends on the change of the transparency of the IGM with respect to the redshift of reionization. We compute $P_{\rm m, x_{\rm HI}}$ by modifying the code available in {\sc 21cmFAST}\citep{2011MNRAS.411..955M}. It is important to highlight that in fact what gets computed from this procedure is the dimensionless power spectrum $\Delta_{\rm m, x_{\rm HI}}$ and hence $\Delta_{\rm m, \psi}$. 


We have not included fluctuations in the low-redshift ionizing background, which will also change the flux power spectrum and have been calculated elsewhere \citep{2014PhRvD..89h3010P,2014MNRAS.442..187G}. To lowest order, the ionizing background effects should be added to the inhomogeneous reionization effects. The ionizing background effect is most prominent at large scales; for example, in the default model of \citet{2014PhRvD..89h3010P} at $z_{\rm obs}=2.3$, the change in power spectrum in the range of scales of interest to us is $\Delta P_{\rm F}^{\rm 3D}(k)/P_{\rm F}^{\rm 3D}(k) \sim 2\Delta b_{\rm HI}/b_{\rm HI} \sim -0.08(k/0.14\,{\rm Mpc}^{-1})^{-1}$ (see Figure 2 of \citealt{2014PhRvD..89h3010P}).\footnote{The $k^{-1}$ dependence at scales small compared to the photon mean free path is the result of the expansion of $S(k)$.} This is similar in magnitude to the reionization effect considered here.

Furthermore, implementation of the effects of He{\,\sc ii} reionization in the IGM \citep{2011MNRAS.410.1096B, 2018arXiv180804367W} may be crucial to correctly compute the impact in the  \lya forest of H{\,\sc i} reionization. Besides, our simulations use simplified models of hydrogen reionization, and the implementation of several key physics will be the scope of future work (e.g. pre-heating before reionization, variations in the thermal and reionization histories). See \S \ref{sec:setup} and \S \ref{sec:summary} for more details.

\section{Simulations}
\label{sec:setup}

To predict the effect of reionization on the \lya forest, we need two types of simulations.

The ``small box'' simulation uses a box size that is smaller than the typical scale of a reionization bubble ($L = 2.55$ Mpc), and can resolve structures down to below the Jeans scale. Its purpose is to determine how the transmitted flux of the \lya forest depends on the reionization redshift $z_{\rm re}$, i.e., to determine the function $\psi(z_{\rm re})$. The small boxes are assumed to reionize instantaneously, and to determine the functional form $\psi(z_{\rm re})$ we run a sequence of boxes with $z_{\rm re}$ stepped over the interesting range (here 6--12). This way we account for correlations between density  and different reionization scenarios. The small boxes have high resolution, since the thermal state of the IGM depends on the way in which small-scale structures are disrupted following reheating \citep{2018MNRAS.474.2173H}. These boxes do not need to self-consistently determine where the ionizing sources are since reionization is externally triggered at a particular redshift. 

In contrast, the ``large box'' simulations are large compared to the scale of reionization bubbles. We choose $L = 300 \ \textup{Mpc}$ such that the simulations have enough statistical power. Their purpose is to predict the spatial structure of the bubbles (reionization) and their correlation with matter. The simulations have a prescription for placing ionizing sources, and a fast approximation scheme to track the fate of ionizing photons. 

It is only with the combination of the two types of boxes that we can compute all of the ingredients in \S\ref{sec:formalism}. Note that these simulations incorporate different physical ingredients and hence we will use two completely separate codes. An important caveat of our simulation strategy is that we miss higher-order correlation functions between the reionization field (on large scales) and the small-scale density field -- due to the small-scale boxes being hydro simulations.

\subsection{Small boxes}
\label{ssec:smallbox}
We follow \cite{2018MNRAS.474.2173H} in the methodology of the small box simulations, but we only use the biggest box size studied in their work. In particular, we use a modified version of {\sc Gadget-2} \citep{2001NewA....6...79S,2005MNRAS.364.1105S}, which is a smoothed particle hydrodynamics (SPH) code. In this modified {\sc Gadget-2} we include the most relevant heating and cooling processes for the gas. {\sc Gadget-2} comes with adiabatic expansion/contraction (including Hubble expansion) and shock heating already implemented. We add Compton heating/cooling for neutral gas (with residual ionization) and Compton cooling for the ionized gas; see Eq.~(17) in \cite{2018MNRAS.474.2173H}. Also for ionized gas we include recombination cooling, photoionization heating, free-free cooling, and He{\sc\,ii} line cooling; see Eqs.~(19--23) in \cite{2018MNRAS.474.2173H}. Following the default treatment in \cite{2018MNRAS.474.2173H}, reionization is treated with an uniform post-reionization temperature of $T_{4,\rm re} = 2$ (or $T_{\rm re} = 2 \times 10^4 \ \textup{K}$) everywhere. Thereafter, photoionization heating is implemented with an energy injection of 4.2 eV per H{\sc\,i} ionization and 7.2 eV per He{\sc\,i} ionization.

Each simulation we used in {\sc Gadget-2} has the same box size, $L = 1728 \, h^{-1} = 2551 \, \textup{kpc}$ and the same number of particles, $2 \times (384)^3$. The dark matter particle mass in the simulations is $9.72 \times 10^3 \ \textup{M}_{\odot}$, while the gas particle mass is $1.81 \times 10^3 \ \textup{M}_\odot$. The only difference among them is when reionization turns on. We simulate eight realizations in order to reduce the statistical error due to the limited box size by a factor of $\sqrt{8}$.

As in \cite{2018MNRAS.474.2173H} we start all the simulations at recombination, $z_{\rm dec} = 1059$ with a modified version of {\sc N-Gen-IC} (the default initial condition generator in {\sc Gadget-2}) to enable streaming velocities between the baryons and dark matter \citep{2010PhRvD..82h3520T}. The boxes are evolved with neutral gas physics until reionization, at which point the temperature is reset to $2\times 10^4$ K, and the box is evolved further with singly ionized (H$^+$/He$^+$) primordial gas physics. There is no He{\sc\,ii} reionization in these simulations. These simulations were tested for convergence in \S 5.3 of \cite{2018MNRAS.474.2173H}, including varying the box size to test the effects of missing large-scale power. The missing variance at large-scales for our small-scale boxes with $L = 2551 \ \textup{kpc}$ is $\sigma^2(z = 2.5) = 0.77$. For comparison, \citet{2018MNRAS.474.2173H} also ran the ``II-C'' boxes with $L = 1275 \ \textup{kpc}$ or $\sigma^2(z = 2.5) = 1.32$. For $z_{\rm re}=7$ vs.\ 8 and 9 vs.\ 8, and for $2.5\le z_{\rm obs}\le 4.0$, $\Delta\ln\tau_1$ changing by $<10\%$ or $<2\sigma$. 

These simulations generate a map of optical depth with arbitrary normalization \citep{2018MNRAS.474.2173H}. The requirement to reproduce the observed mean flux sets the normalization. Furthermore, the required normalization is reported in the form of $\tau_1$, and hence we use $\tau_1$ to show the dependence of the transmission of the \lya forest with the redshift of reionization.

All of the small box simulations were done on the Ruby cluster at the Ohio Supercomputer Center \citep{Ruby2015}.

\subsection{Large boxes}

We include the physics of reionization-bubble-scale by using semi-numerical simulations, specifically we use {\sc 21cmFAST} \mbox{\citep{2011MNRAS.411..955M}}. We utilize all the default parameters in {\sc 21cmFAST} with the exceptions of a few parameters that we change for different alternatives scenarios in order to examine different reionization histories, and the use of our chosen background cosmology. The box size for each one of these simulations is $L=300 \ \textup{Mpc}$. We run eight realizations for each of the three different reionization scenarios with the same physical setup but with different random seeds to diminish the simulation variance. Furthermore, we generate snapshots that track -- among other parameters -- the density, neutral fraction of hydrogen and the 21 cm temperature fields for $34.70\ge z\ge 5.90$, with a step size of 2\%, i.e., $z_{<} = (z_{>} + 1) / 1.02 - 1$. Therefore, in total we have $84$ snapshots (each with 3 dependent variables) for the computation of the cross-power spectrum of matter and neutral fraction, and the 21 cm quadrupole per realization. In order to compute the cross-power spectra we only modify ``delta\_ps'' in {\sc 21cmFAST}, such that it can take both a density and neutral fraction snapshot. Furthermore, to be able to compute the 21 cm quadrupole we include the ability to compute $P_T(k,\mu)$ instead of $P_T(k)$.

To explore different reionization histories we change the number of ionizing photons escaping into the IGM per baryon in collapsed structures, i.e. \texttt{HII\_EFF\_FACTOR} in {\sc 21cmFAST}, which represents the ionizing efficiency
\begin{eqnarray}
\label{eq:reiohisto}
\zeta = N_{\gamma/{\rm b}} \, f_{\rm esc} f_* \, f_{\rm b} \, \textup{,}
\end{eqnarray}
where $f_{\rm b}$ is the baryon fraction of a halo in units of the cosmic mean value $\Omega_{\rm b}/\Omega_{\rm m}$, $f_*$ is the fraction of baryons from the halos that form  stars, $N_{\gamma/{\rm b}}$ corresponds to the number of ionizing photons produced per stellar baryon, and $f_{\rm esc}$ is the fraction of produced ionizing photons which escape into the IGM (see \citealt{2018PhR...780....1D} for more details regarding the physics behind Eq.~(\ref{eq:reiohisto})). In contrast to our approach here, \cite{2018arXiv181011683O} used both $\zeta$ and the minimum halo mass for producing ionizing photons as free parameters for their hybrid simulations. 

\begin{figure}
\includegraphics[width=\columnwidth]{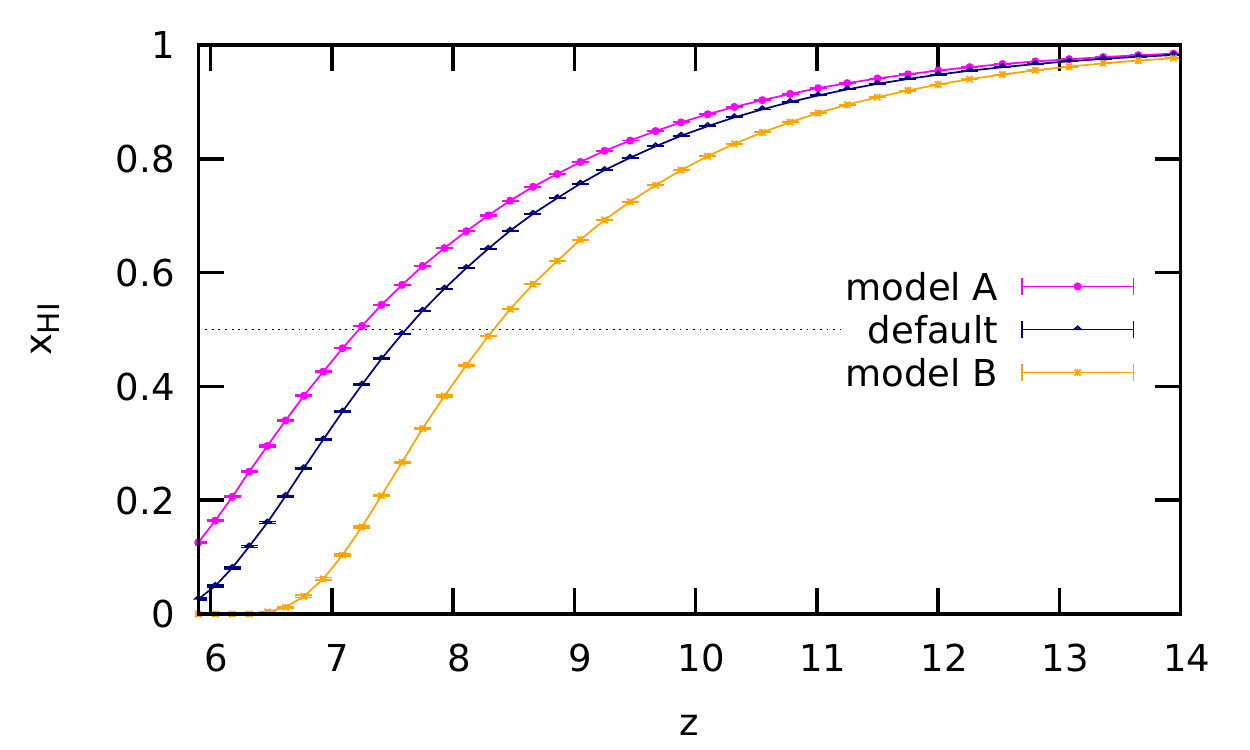}
    \caption{Neutral fraction history for the different models used in our large boxes. Our models differ only by the number of ionizing photons escaping into the IGM per baryon, with model A having the latest reionization, and model B having the earliest reionization.}
    \label{fig:neutralfraction}
\end{figure}

We define a \textit{default} model with optical depth to reionization equal to the current best fit value of the recent final results of \textit{Planck} 2018 `$TT + TE + EE +lowE + {\rm lensing}$'  \citep{2018arXiv180706209P}, i.e. $\tau = 0.054$, which corresponds to an ionizing efficiency of $\zeta = 25 $.\footnote{The main difference between the \textit{Planck} 2015 and 2018 results is the value of the optical depth which in 2015 was $\tau = 0.066$. We keep all other cosmological parameters equal to the cosmology from \textit{Planck} 2015 used throughout this paper.} In our default model this optical depth corresponds to a redshift of reionization of $z_{\rm re} = 7.61$, which corresponds to when the mean neutral hydrogen fraction is $0.5$. For comparison purposes we chose alternative models with later (``model A'') and earlier (``model B'') reionization than our default model. We select these alternative scenarios so that they encompass the range of models consistent with the new Planck results and other reionization probes at $\sim \pm 1 \sigma$ (see \citealt{2015ApJ...811..140B} for a compact list of these probes). We choose our model A such that the volume weighted neutral faction at $z = 5.9$ approximately matches the $1\sigma$ upper limit extracted from the dark pixel measurements reported in \mbox{\cite{2015MNRAS.447..499M}}, $\bar{x}_{\rm HI} \leq 0.11$.\footnote{We note that our model A might be considered as conservative in light of recent high redshift constraints of the hydrogen neutral fraction, e.g., $x_{\rm HI} = 0.88$ at $z = 7.6$ \citep{2019arXiv190109001H} and $x_{\rm HI} > 0.76$ at $z \sim 8$ \citep{2019arXiv190111045M}.} This corresponds to roughly $\zeta = 20.9$, and by construction the reionization process has not been completed by the end of our large simulations.\footnote{The lower $\tau$ from \textit{Planck} was reported too late to change the small-box simulations to run to lower $z$.} Model B has $\zeta = 35$ with $z_{\rm re} = 8.35$, which is $1\sigma$ away from the Planck optical depth.  See Fig.~\ref{fig:neutralfraction} and Table \ref{tab:models} for a description of the models.

\begin{table}
\caption{Summary of the different reionization models. Note that since model A has not finished the reionization process by redshift 5.90, its optical depth is not as accurate as the other models. Furthermore, since the volume weighted neutral fraction differs from the mass weighted one at the end of the reionization process, the optical depths are approximations.}
\label{tab:models}
\begin{tabular}{lccc}
\hline \hline
Reionization model & $\tau$ & $z_{\rm re}$ & $\zeta$\\
\hline
A: later reionization & 0.0512 & 7.22  & 20.9\\
Default: Planck 2018 reionization & 0.0548 & 7.61 & 25\\
B: earlier reionization & 0.0615 & 8.34 & 35\\
\hline \hline
\end{tabular}
\end{table}

\section{Assessment of contamination}
\label{sec:contamination}

The goal of this work is to quantify the effect of inhomogeneous reionization in the \lya flux. In principle to estimate the contamination to the signal we simply require both terms in the RHS of equation (\ref{eq:pof}), i.e. the linear theory term and the non-linear extension. 

\subsection{Linear power spectrum (3D)}
\label{ssec:linear_power}
We start by computing the linear term. We simplify this process by assuming that $\mu = 0$, i.e. perpendicular to the line of sight so that one can ignore redshift space distortions. We compute the matter power spectrum from the density boxes generated in \texttt{21cmFAST} with our chosen cosmology. For the sake of illustrating the contamination we choose a redshift of observation of $2.5$ and a wavenumber of $0.14 \ \textup{Mpc}^{-1}$, both values are typical of \lya measurements. Then, we have  
\begin{eqnarray}
\label{eq:comp1}
P^{\rm 3D}_{\rm F} (z_{\rm obs} = 2.5, k = 0.14 \ \textup{Mpc}^{-1}) = 995.3 b^2_{\rm F} \, (z_{\rm obs}) \ \textup{Mpc}^3 \, \textup{.}
\end{eqnarray}

We extract the \lya flux bias from Table 1 in \cite{2011MNRAS.415.2257M}. We summarize the relevant values in Table \ref{tab:bias}. Since we are using the results from \cite{2011MNRAS.415.2257M} we will be consistent with their assumption that the redshift distortion parameter is equal to unity, i.e. $\beta_{\rm F} = 1$ throughout our work. 

\subsection{Computation of $\psi(z_{\rm re})$}
\label{ssec:get_psi}

We use our smaller simulations to calculate how the transparency of the \lya forest depends on when reionization occurs. Our small-box simulations (\S\ref{ssec:smallbox}) compute  \lya transmission by the procedure described in detail in \S4.4 of \cite{2018MNRAS.474.2173H}. Here we briefly describe the strategy used, and the methodology for extracting $\psi$ from our simulations. 

Each run of our small-scale simulations generates eight snapshots of optical depth maps starting at $z= 5.5$. They are separated by decrements of $\Delta z = 0.5$. We run our small simulations for seven different redshifts of reionization, $z_{\rm re} = \{6,7,8,9,10,11,12\}$. We chose $\bar{z}_{\rm re} = 8$ as our reference redshift of reionization, although due to the definition of $\psi$ as a change in $\ln\tau_1$, the choice of reference is simply a constant offset in $\psi$ with no effect on $P_{m,\psi}(k)$.

Once all the relevant snapshots have been generated, one can finally compute how the transparency of the \lya forest depends on the redshift of reionization. We obtain $\psi$ from Eq.~(\ref{eq:psire}) by comparing the $\tau_1$ from Simulation B with redshift of reionization $z_{\rm re}$ to the $\tau_1$ obtained for Simulation A with redshift of reionization $\bar{z}_{\rm re} = 8$. We present our results for $\psi$ and its redshift dependence in Table \ref{tab:finalsimu}. In what follows, we linearly interpolate $\psi$ between the redshifts in the table.  


\begin{table}
\centering
\caption{Bias factors: Flux bias $b_{\rm F}$, radiation bias $b_{\rm \Gamma}$ and ratio of $b_{\rm \Gamma}/b_{\rm F}$  at different redshifts.}
\label{tab:bias}
\begin{tabular}{cccc}
\hline\hline
$z$ & $b_{\rm F}$ & $b_{\rm \Gamma}$ & $b_{\rm \Gamma}/b_{\rm F}$ \\
\hline
2.0 & -0.12 & -0.084 & 0.70 \\
2.5 & -0.18 & -0.146 & 0.81 \\
3.0 & -0.27 & -0.237 & 0.88 \\
3.5 & -0.37 & -0.356 & 0.96 \\
4.0 & -0.55 & -0.505 & 0.92 \\
\hline\hline
\end{tabular}
\end{table}

\begin{table*}
	\centering
	\caption{Results for the small-scale simulations. Transparency variations in the IGM, $\psi = \Delta \ln \tau_1 ({\rm Sim. \, B})/\tau_1 ({\rm Sim. \, A})$. The number in square brackets is the redshift at which reionization turns on. Note that $\Delta \ln \tau_1$ is negative if simulation A is more transparent than simulation B.}
	\label{tab:finalsimu}
	\begin{tabular}{cccccccc}
		\hline\hline
		Box size & Sim. A & Sim. B & \multicolumn{5}{c}{$10^5 \times \Delta \ln \tau_1$}\\
        $[\textup{ckpc}]$ & {} & {} & $z=2.0$ & $z =2.5$ & $z = 3.0$ & $z=3.5$ & $z=4.0$\\
		\hline
		2551 & [8] & [6] & 6064 $\pm$ 553 & 6910 $\pm$ 895 & 9772 $\pm$ 973 & 14616 $\pm$ 1068 & 22522 $\pm$ 1476\\
        2551 & [8] & [7] & 2112 $\pm$ 235 & 2399 $\pm$ 367 & 3612 $\pm$ 436 & 5769 $\pm$ 443 & 9275 $\pm$ 550\\
		2551 & [8] & [9] & -878.8 $\pm$ 166 & -649.6 $\pm$ 252 & -1083 $\pm$ 315 & -2548 $\pm$ 299 & -4682 $\pm$ 379\\
        2551 & [8] & [10] & -1300 $\pm$ 303 & -407.5 $\pm$ 406 & -847.8 $\pm$ 560 & -2950 $\pm$ 617 & -6344 $\pm$ 659\\ 
        2551 & [8] & [11] & -1425 $\pm$ 413 & -52.51 $\pm$ 582  & -318.9 $\pm$ 724 & -2626 $\pm$ 780 & -6610 $\pm$ 899 \\
        2551 & [8] & [12] & -1552 $\pm$ 482 & 199.7 $\pm$ 674 & 283.5 $\pm$ 850 & -2128 $\pm$ 912 & -6400 $\pm$ 1007\\
		\hline\hline
	\end{tabular}
\end{table*}

\subsection{Cross-power spectrum of matter and $\psi$}
\label{ssec:p_mpsi}
\begin{figure}
	\includegraphics[width=\columnwidth]{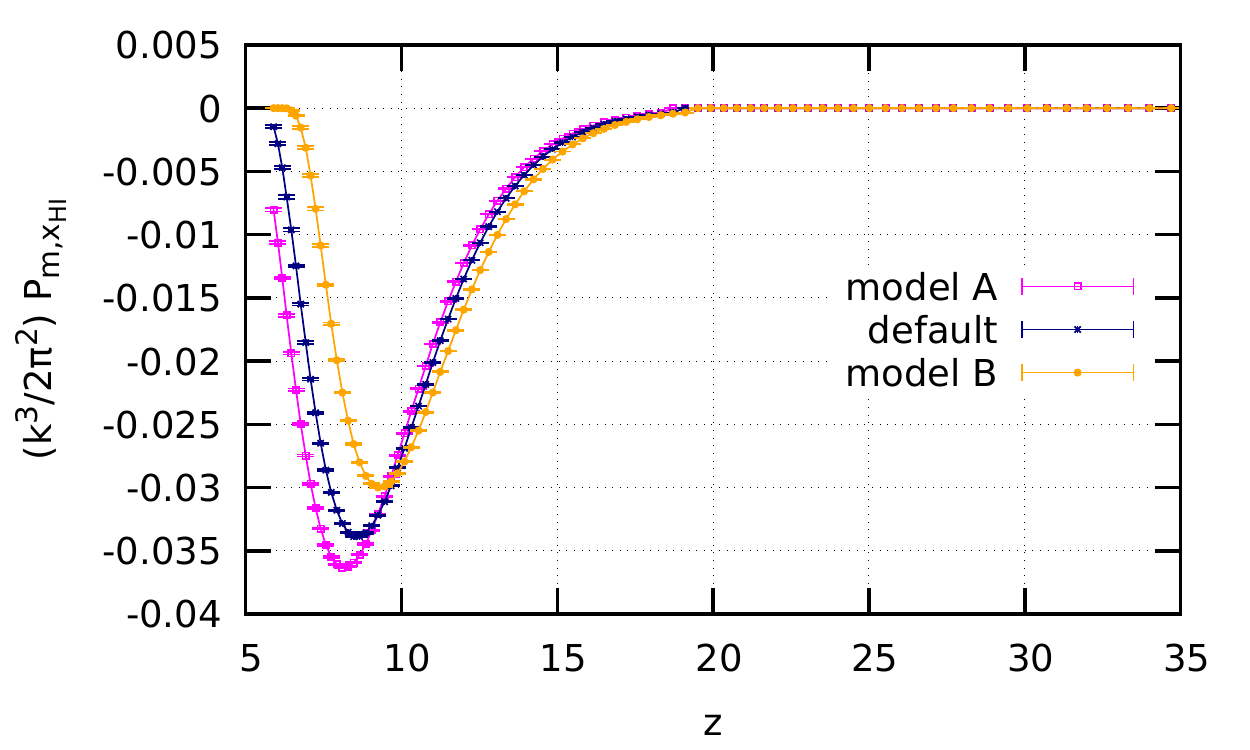}
    \caption{Dimensionless cross-power spectrum of matter and neutral hydrogen fraction as a function of redshift for our different models of reionization history. All cross-power spectra have been evaluated at wavenumber $k = 0.14 \ \textup{Mpc}^{-1}$.}
    \label{fig:deltamxh}
\end{figure}

In Eq.~(\ref{eq:crossps}) we accounted for the inhomogeneous nature of reionization by including the cross-power spectrum of matter and hydrogen neutral fraction. We obtain this cross-power spectrum from our modified version of \texttt{21cmFAST}. We plot the dimensionless cross-power spectrum of matter and neutral fraction for the different models in Fig.~\ref{fig:deltamxh}. Note that $P_{\rm m, x_{HI}}$ is not only negative, because overdense regions will ionize first resulting in an anti-correlation between matter density and neutral fraction. The absolute value peaks around the redshift of reionization, because both before and after reionization the perturbations in $x_{\rm HI}$ go to zero.

We determine the impact of inhomogeneous reionization on the \lya flux from both our calculations of how the transparency of the IGM depends on the redshift of reionization \S \ref{ssec:get_psi}, and the cross-correlation of mater and the neutral hydrogen fraction. Using Eq.~(\ref{eq:crossps}) evaluated at $k = 0.14 \ \textup{Mpc}^{-1}$ and observing at redshift $2.5$ for our default model, we obtain $P_{\rm m, \psi} = -26.7 b_{\rm F}(z_{\rm obs}) b_{\rm \Gamma} (z_{\rm obs}) \ \textup{Mpc}^3$. Therefore we see that the contamination to the linear term in Eq.~(\ref{eq:comp1}), ignoring the bias factors, is approximately $5.36 \%$. We ignored the bias ratio to quantify this change, however as seen in Table \ref{tab:bias} both bias factors can significantly affect the deviation. We tabulated the deviation of the \lya power spectra, taking into account the role of the bias factors, for wavenumber $0.14 \ \textup{Mpc}^{-1}$ and different redshifts of observation in Table \ref{tab:result}. We obtained the radiation bias directly from our \texttt{GADGET 2} simulations by computing $b_{\rm \Gamma} = \partial \ln \bar{F}/\partial \ln \tau_1$\footnote{Note that the bias factors obtained from our simulations seem to be slightly higher than the ones reported in \cite{2018MNRAS.474.2173H}; nevertheless, we consider these good enough for a first estimation of the effects of inhomogeneous reionization in the \lya flux.} (see Table \ref{tab:bias}). 

\begin{table*}
\centering
\caption{Percentage deviation of the 3D and 1D \lya power spectrum due to patchy reionization for the different reionization models considered. Here we have used $k = 0.14 \ \textup{Mpc}^{-1}$, and we included the bias ratio.}
\label{tab:result}
\begin{tabular}{cccccc}
\hline\hline
Simulation & \multicolumn{5}{c}{Ratio of $2 \times (b_{\Gamma}/b_{\rm F}) \times P_{\rm m, \psi}/P_{\rm m} \times 100\% $ with $k = 0.14 \ \textup{Mpc}^{-1}$ }\\ 
{} & $z_{\rm obs} = 2.0$ & $z_{\rm obs} = 2.5$ & $z_{\rm obs} = 3.0$ & $z_{\rm obs} = 3.5$ & $z_{\rm obs} = 4.0$ \\
\hline
Model A 3D & 4.09 $\pm$ 0.47 & 5.56 $\pm$ 0.98 & 9.85 $\pm$ 1.45 & 19.6 $\pm$ 1.87 & 35.9 $\pm$ 2.63 \\
Default 3D & 3.39 $\pm$ 0.42 & 4.35 $\pm$ 0.87 & 7.82 $\pm$ 1.32 & 16.4 $\pm$ 1.69 & 31.0 $\pm$ 2.31 \\
Model B 3D & 1.96 $\pm$ 0.32 & 1.84 $\pm$ 0.66 & 3.45 $\pm$ 1.05 & 9.10 $\pm$ 1.34 & 19.3 $\pm$ 1.73 \\
\hline 
Model A 1D & 0.75 $\pm$ 0.08 & 0.92 $\pm$ 0.17 & 1.81 $\pm$ 0.28 & 3.63 $\pm$ 0.34 & 6.53 $\pm$ 0.42 \\
Default 1D & 0.61 $\pm$ 0.08 & 0.68 $\pm$ 0.16 & 1.36 $\pm$ 0.26 & 2.92 $\pm$ 0.31 & 5.45 $\pm$ 0.37 \\
Model B 1D & 0.35 $\pm$ 0.06 & 0.26 $\pm$ 0.12 & 0.52 $\pm$ 0.21 & 1.52 $\pm$ 0.26 & 3.27 $\pm$ 0.29 \\
\hline\hline
\end{tabular}
\end{table*}

\begin{figure}
    \centering
    \begin{minipage}{\linewidth}
\centering
\includegraphics[width=\columnwidth]{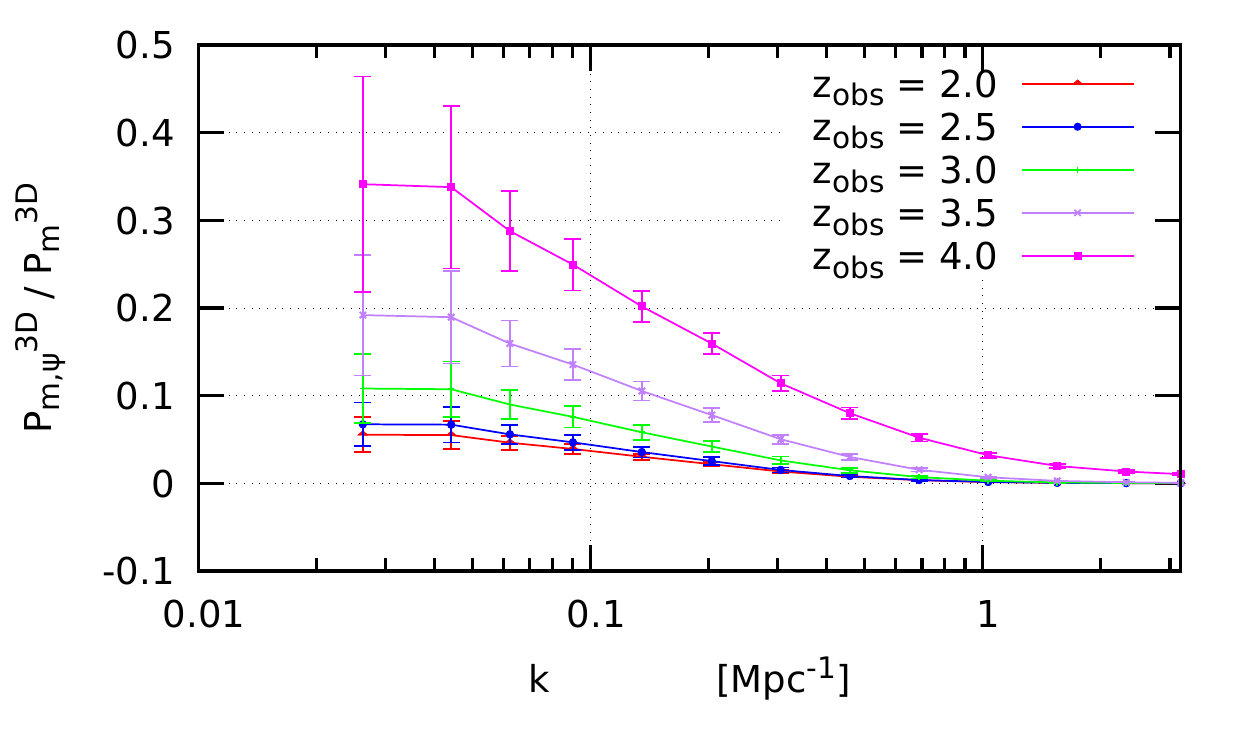}
{\textbf{(a).} Model A: Later reionization.}
\end{minipage}
\begin{minipage}{\linewidth}
\centering
\includegraphics[width=\columnwidth]{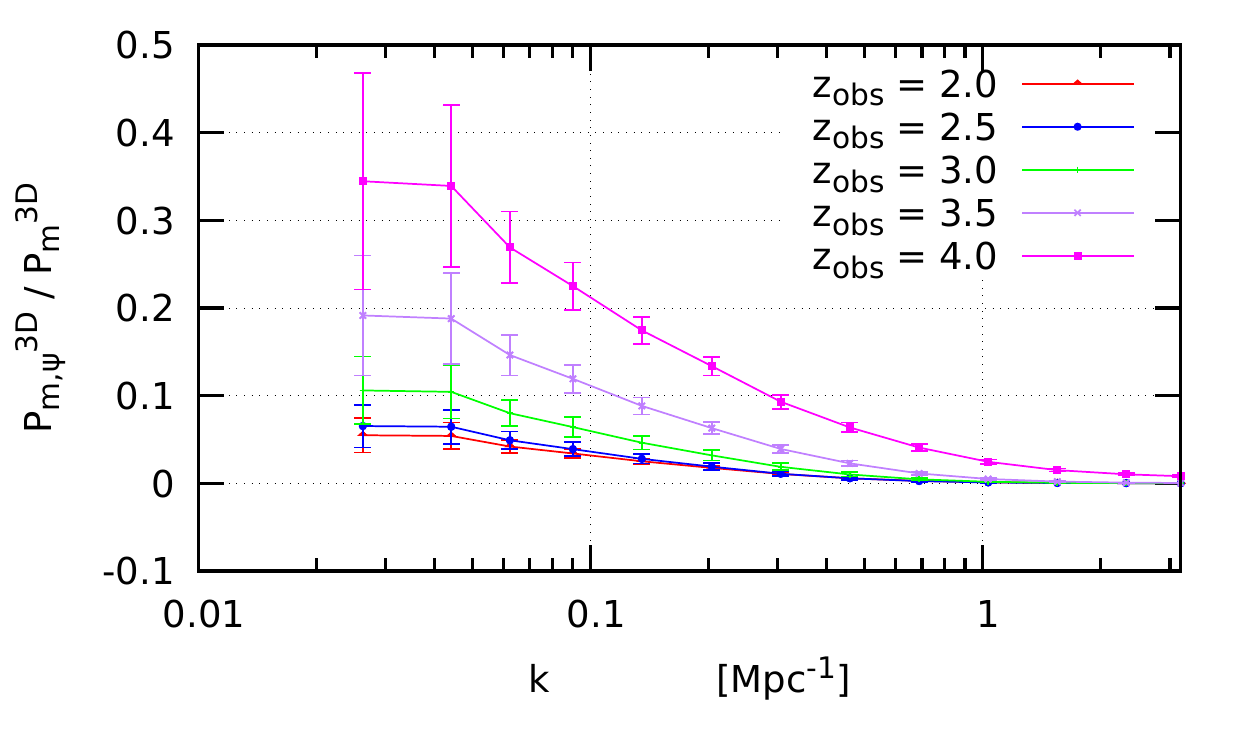}
{\textbf{(b).} Default model: \textit{Planck} 2018 reionization.}
\end{minipage}
\begin{minipage}{\linewidth}
\centering
\includegraphics[width=\columnwidth]{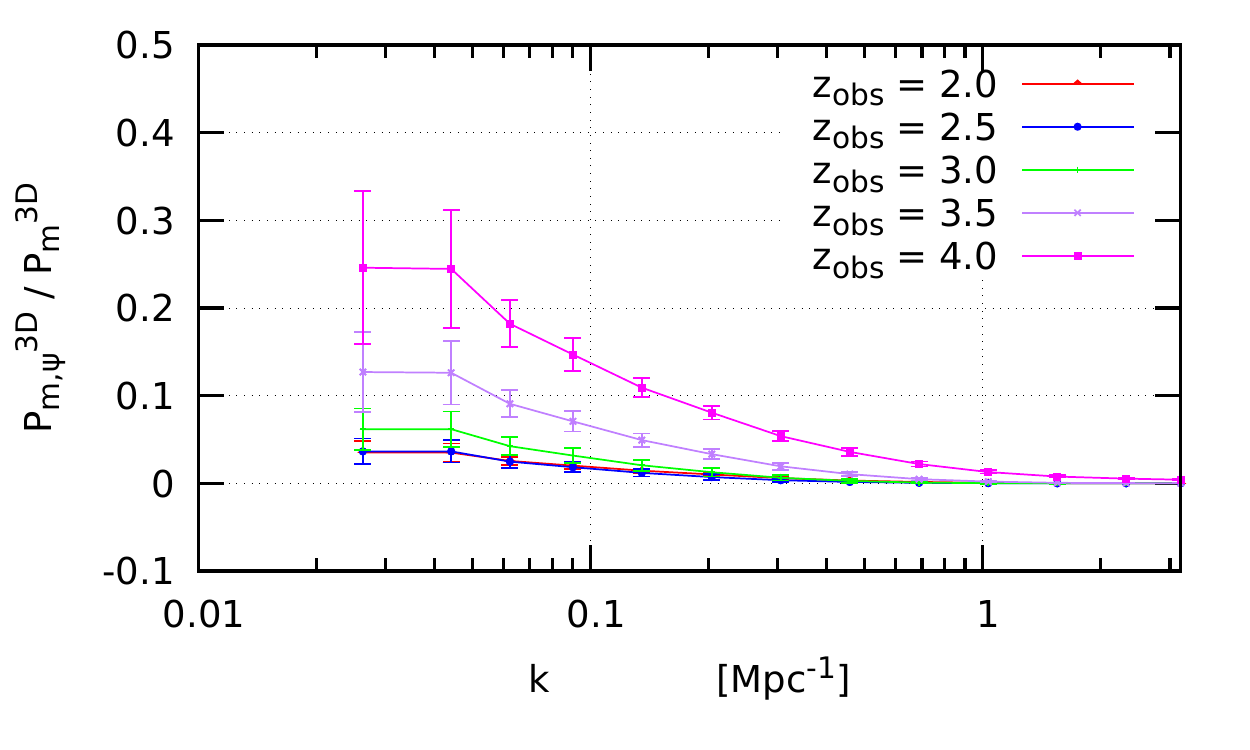}
{\textbf{(c).} Model B: Earlier reionization.}
\end{minipage}
    \caption{Comparison for the different reionization models between the dimensionless cross-power spectrum of $\psi$ and matter, and the dimensionless power spectrum of matter as a function of wavenumber for fixed redshift of observation.}
    \label{fig:deltacomp}
\end{figure}

In Fig.~\ref{fig:deltacomp} we plot a comparison of the strength of the effect by looking at the ratio of the 3D dimensionless cross-power spectrum of the reionization effect and the mass auto-power spectrum as a function of wavenumber. The hierarchical structure in redshift of observation for all the models is anticipated, independent of the scale. The higher the redshift, the larger the effect of thermal relics from reionization on the IGM. That is because there is not enough time to relax onto the usual temperature-density relation. In addition, the ratio is larger at large scales, which is expected since it couples to the reionization bubble scales. 

The error bars in Fig.~\ref{fig:deltacomp} and Table \ref{tab:result} were computed by combining the errors from our large-box and small-box simulations. Because both simulations are independent of each other -- i.e. the error sources are orthogonal -- and both enter ``linearly'' into the computation of $P_{{\rm m}, \psi}$, one can average over the realizations of the small-box simulations and then compute the sample variance on the mean for the large-box simulations, and vice-versa. Finally, the complete error bars are computed by adding the two variances in quadrature, i.e.
\begin{eqnarray} 
\frac{1}{64} \sum_{ij} f(\psi_i, {\rm F}_j) = f\left( \frac{1}{8} \sum_{i = 0}^{N_i = 8} \psi_i, \frac{1}{8} \sum_{j=0}^{N_j=8} {\rm F}_j \right) \, \textup{,}\\
\Rightarrow {\rm error}  =  \sqrt{SV\left({\rm F}_j,\frac{1}{8} \sum_{i=0}^{N_i = 8} \psi_i \right)^2 +  SV\left(\psi_i,\frac{1}{8} \sum_{j=0}^{N_{j}=8} {\rm F}_j \right)^2} \, \textup{,}
\end{eqnarray}
where $f$ stands for our algorithm to compute $\Delta P/P$, SV stands for sample variance on the mean, the $\psi_i$ represent the small-scale realizations and ${\rm F}_j$ are the large-scale realizations. 

\subsection{Linear power spectrum (1D)}
\label{ssec:linear_1D}
Once we have the 3D power spectrum (the cross-correlation between different skewers) we can compute the 1D \lya power spectrum (the cross-correlation between pixels of the same skewer) which is given by averaging over the perpendicular direction to the line of sight, i.e. 
\begin{eqnarray}
P^{\rm 1D}_{\rm F} (k,z_{\rm obs})  & = & \int_0^\infty\frac{dk_{\perp}}{2 \pi} k_\perp  P^{\rm 3D}_{\rm F}(k, z_{\rm obs})  \nonumber\\
& = & b_{\rm F}^2  \int_0^\infty \frac{dk_{\perp}}{2 \pi} k_\perp (1 + \mu^2)^2  P^{\rm 3D}_{m}(k,z_{\rm obs}) \nonumber \\
&&+ 2 \, b_{\rm F} \, b_{\rm \Gamma}  \int_0^\infty \frac{dk_{\perp}}{2 \pi} k_\perp (1 + \mu^2)  P^{\rm 3D}_{\rm m,\psi} (k, z_{\rm obs}) \nonumber \\
& = & b_{\rm F}^2 P^{\rm 1D}_{\rm m} (k,z_{\rm obs}) + 2 \, b_{\rm F} \, b_{\rm \Gamma} P^{\rm 1D}_{\rm m, \psi} (k, z_{\rm obs}) \, \textup{.}
\end{eqnarray}
Since our \texttt{21cmFAST} simulations use a $k_{\rm max} \approx 3.20 \ \textup{Mpc}^{-1}$, we set this wavenumber as the upper limit of the integrals over the perpendicular direction.

\begin{figure}
    \centering
    \begin{minipage}{\linewidth}
\centering
\includegraphics[width=\columnwidth]{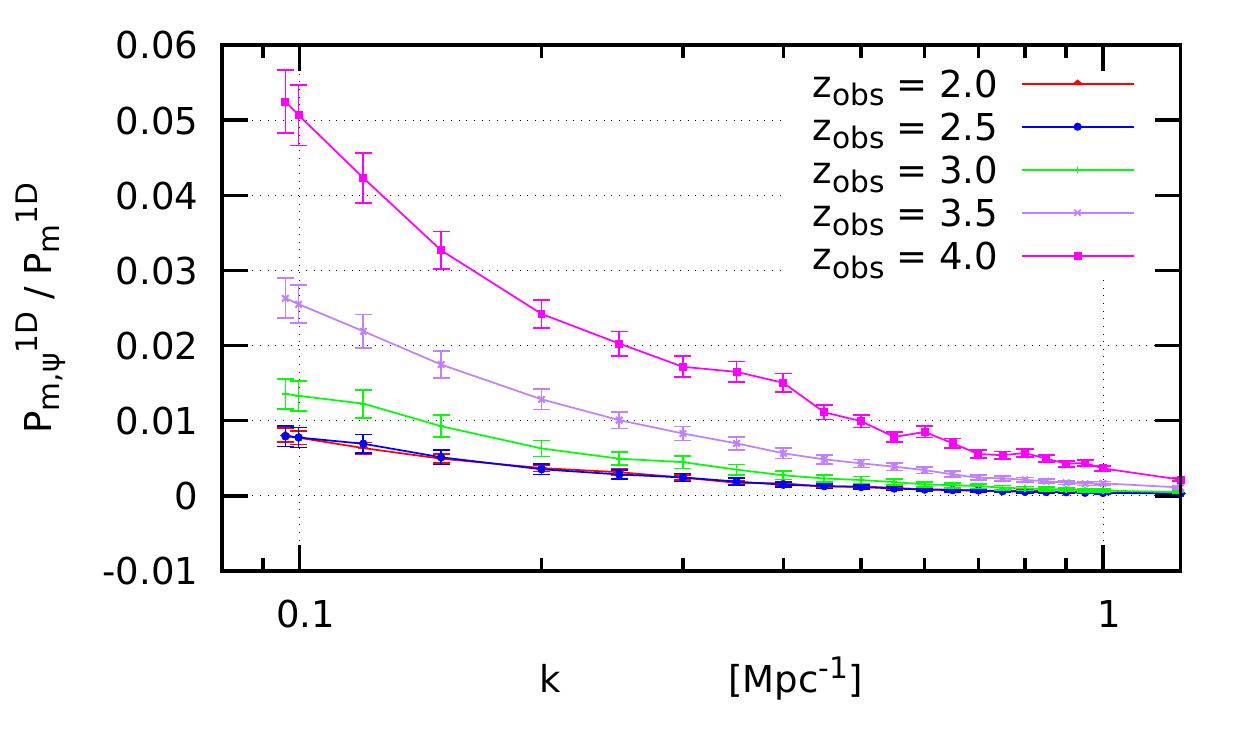}
{\textbf{(a).} Model A: Later reionization.}
\end{minipage}
\begin{minipage}{\linewidth}
\centering
\includegraphics[width=\columnwidth]{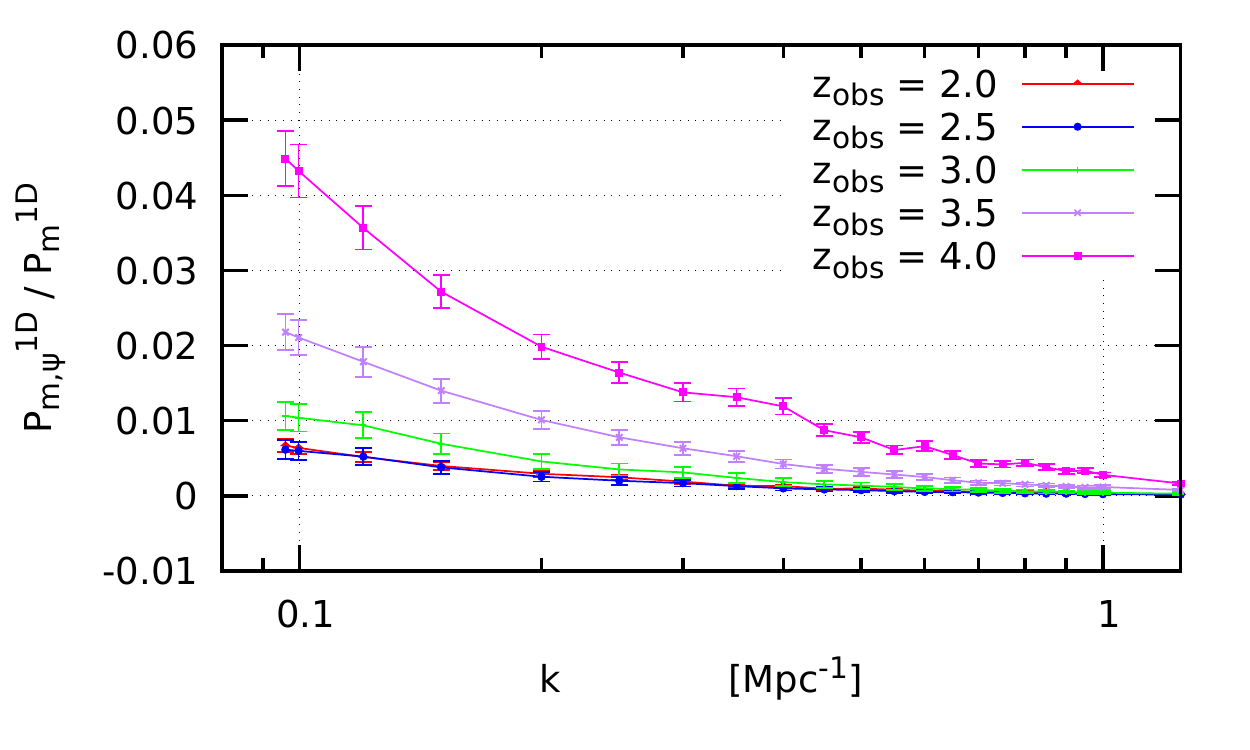}
{\textbf{(b).} Default model: \textit{Planck} 2018 reionization.}
\end{minipage}
\begin{minipage}{\linewidth}
\centering
\includegraphics[width=\columnwidth]{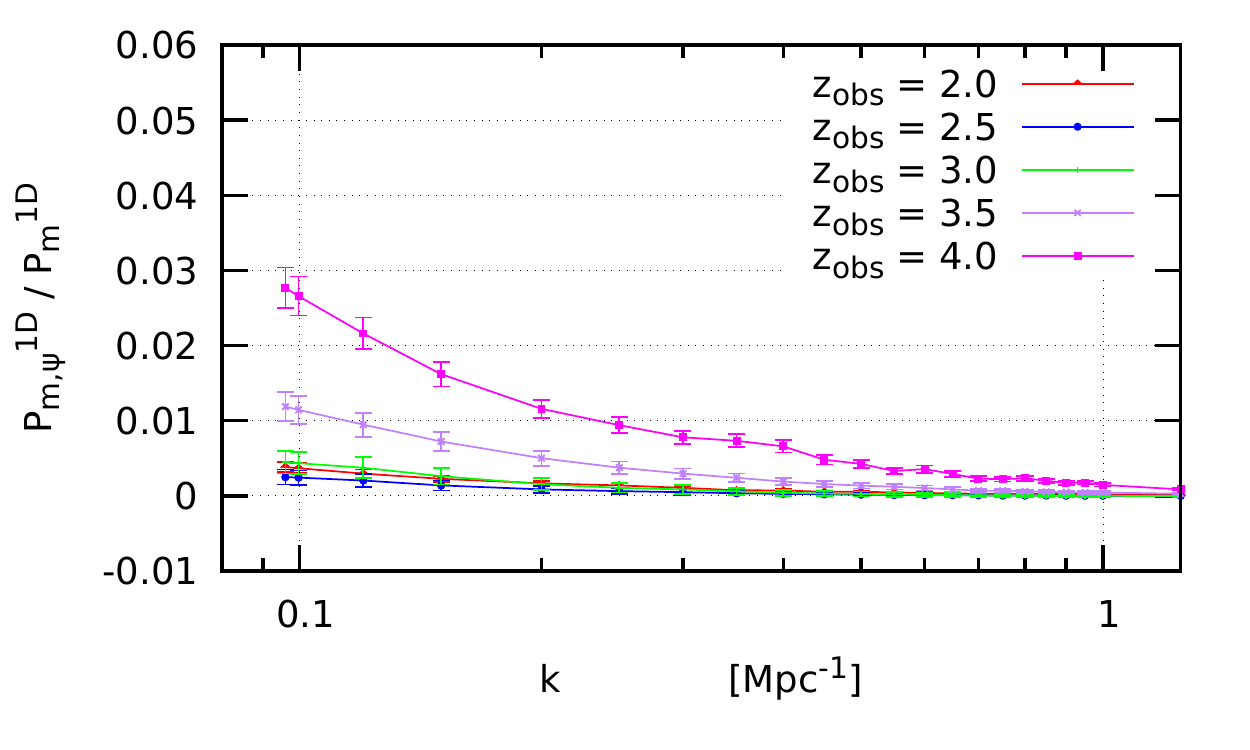}
{\textbf{(c).} Model B: Earlier reionization.}
\end{minipage}
    \caption{Comparison between the 1D dimensionless cross-power spectrum of $\psi$ and matter, and the 1D dimensionless power spectrum of matter as a function of wavenumber for fixed redshift of observation.}
    \label{fig:1Dcomp}
\end{figure}

We report the percentage of the deviation, i.e. the ratio of $2(b_{\rm F}/b_{\rm \Gamma}) P^{\rm 1D}_{\rm m, \psi}/P^{\rm 1D}_{\rm m} \times 100\%$ from our simulations in Table \ref{tab:result}. Furthermore, we plot the ratio of $P^{\rm 1D}_{\rm m, \psi}/P^{\rm 1D}_{\rm F}$ as a function of wavenumber for the different redshifts of observation that we explored in Fig.~\ref{fig:1Dcomp}. We use the 1D \lya power spectrum from \texttt{BOSS} \citep{2013A&A...559A..85P}. From Fig.~\ref{fig:1Dcomp} and Fig~\ref{fig:deltacomp} we confirm that the effect of reionization is stronger for the \lya 3D power spectrum than for the \lya 1D power spectrum, as one could have expected due to the integration smoothing the deviation. Moreover, the 1D ratio is also consistent with the redshift hierarchical structure (within the error bars). The error bars shown in Fig.~\ref{fig:1Dcomp} have been computed in the same way we described in \S \ref{ssec:p_mpsi}, and hence we have ignored the error bars from the \texttt{BOSS} data, since they are small compared to the uncertainties in the reionization models. We note that our results for the effect of inhomogeneous reionization in the 1D \lya forest power spectrum at $z_{\rm obs} = 4.0$ are slightly smaller than the ones reported in \cite{2018arXiv181011683O} -- see their Fig.~(7) -- but the trend at large scales are consistent and the differences in numerical values can be attributed to model dependence. 

Finally, as could have been foreseen, the imprint from reionization affects the \lya forest the most for our model of late reionization (model A) since the IGM has the least time to ``thermalize'' for $z_{\rm re} \approx 7.25$. In contrast, the early reionization model suffers the lowest effect due to having more time to dilute the thermal relics, and achieve the usual temperature-density relation. Furthermore, the overall magnitude of the effect is surprisingly strong. Hence, it will be crucial to correctly account for this systematic signal in both near-term and future \lya forest surveys. 

\section{Constraining the effect with 21 cm cosmology}
\label{sec:21cmcross}
In the previous section we described a potentially very significant systematic for \lya forest measurements, one that is particularly important for later reionization scenarios. In this section we will illustrate how this systematic could become an interesting link between \lya and 21 cm cosmology.

In this section we show how non-parametric mitigation of the thermal imprint of reionization in the \lya forest is possible through the use of the ``linear $\mu_{\boldsymbol k}$-decomposition'' scheme \citep{2005ApJ...624L..65B}. Unfortunately, as we will illustrate, non-linear effects are significant and hence in future work we will continue our efforts of non-parametric mitigation by employing the ``quasi-linear $\mu_{\boldsymbol k}$-decomposition'' \mbox{\citep{2012MNRAS.422..926M}}.

Under the linear $\mu_{\boldsymbol k}$-decomposition the 3D 21 cm power spectrum is given by \mbox{\citep{2012MNRAS.422..926M}}
\begin{eqnarray}
\label{eq:quasi}
P^{\rm 3D}_{T} (\boldsymbol{k})  =  P_{\mu^0}(k) + P_{\mu^2}(k)\,\mu^2 + P_{\mu^4}(k) \,\mu^4 \, \textup{,}
\end{eqnarray}
where each $P_{\mu} (k) = \langle P_{\mu} (\boldsymbol{k}) \rangle$ is angle-averaged over constant k-shells. Moreover, in the limit of $T_{\rm s} \gg T_{\rm CMB}$, which corresponds to the range where $P_{m, x_{\rm HI}}$ is important (see Fig.~\ref{fig:deltamxh}), Eq.~(\ref{eq:quasi}) becomes
\begin{eqnarray}
\label{eq:pmu0}
P_{\mu^0} (k,z) &=&  (\delta \bar{T}_{\rm b}(z))^2 \, P_{\delta_{\rho_{\rm HI}}} (k,z) \,  \textup{,}\\
\label{eq:pmu2}
P_{\mu^2} (k,z) &=&  2 (\delta \bar{T}_{\rm b}(z))^2 \, P_{\delta_{\rho_{\rm HI}}, \delta_{\rho_{\rm H}}} (k,z) \ \ \, \, \ \  \textup{and} \\
\label{eq:pmu4}
P_{\mu^4} (k,z)  &=&  (\delta \bar{T}_{\rm b}(z))^2 \, P_{\delta_{\rho_{\rm H}}} (k,z) \,  \textup{,}
\end{eqnarray}
where $\delta_{\rho_{\rm HI}} = \delta_{\rho_{\rm H}} + \delta_{x_{\rm HI}} + \delta_{x_{\rm HI}}\delta_{\rho_{\rm H}}$ and $\delta \bar{T}_{\rm b}$ is the mean of the brightness temperature. In the limit of $T_{\rm s} \gg T_{\rm CMB}$ we have 
\begin{eqnarray}
\label{eq:21cmmean}
\delta \bar{T}_{\rm b}(z) & = & \nonumber \frac{3c^3 A_{10} T_{*} \bar{n}_{\rm HI}(z)}{32 \pi \nu^3_{\rm hf} (1+z) H(z)} \\
& \approx & 26.6 \bar{x}_{\rm HI} \left(\frac{\Omega_{b}h^2}{0.0223}\right) \left( \frac{\Omega_m h^2}{0.1417} \frac{1 + z}{10} \right)^{1/2} \textup{mK} \, \textup{,}
\end{eqnarray}
where $A_{10}$ is the Einstein coefficient of the hyperfine transition, $T_{*}$ is the $21$ hyperfine transition in temperature units and $\nu_{\rm hf}$ is the $21$ cm frequency.  

Rewriting Eq.~(\ref{eq:quasi}) into $\ell$-multipoles we have
\begin{eqnarray}
\label{eq:multip}
P^{\rm 3D}_{T} (\boldsymbol{k},z) & = & P_{\mu^0} + \frac{P_{\mu^2}}{3} + \frac{P_{\mu^4}}{5} +\frac{2}{3} \left( P_{\mu^2} + \frac{6}{7} P_{\mu^4} \right) L_{2} (\mu)  \nonumber \\
&& +  \frac{8}{35} P_{\mu^4} L_{4} (\mu) \, \textup{,}
\end{eqnarray}
where $L_{\ell}$ are the Legendre polynomials. Note that in our notation the quadrupole term can be expanded as
\begin{eqnarray}
\label{eq:THEQUAD}
P^{\ell = 2}_{T} = \frac{2}{3}(\delta \bar{T}_{\rm b})^2\left( \frac{2}{\bar{x}_{\rm HI}} P_{m, x_{\rm HI}} + \frac{20}{7} P_m  \right) \, \textup{,}
\end{eqnarray}
where we have taken advantage of the fact that hydrogen traces the matter distribution, and the extra factor of neutral hydrogen fraction comes from the way we have defined our perturbations on the neutral fraction, i.e. $\delta_{x_{\rm HI}} = x_{\rm HI} - \bar{x}_{\rm HI}$. 

Hence the cross-power spectrum of matter and fraction of neutral hydrogen atoms is related through Eq.~(\ref{eq:THEQUAD}) to the $\ell = 2$ multipole component of the $21$ cm power spectrum. Therefore with a measurement of the $21$ cm power spectrum one can in principle constrain the cross-power responsible for the systematic imprint in the \lya forest.

The first step for our mitigation scheme to be successful is to be able to reproduce the 21 cm power spectrum. As was quantified in Fig.~10 of \cite{2012MNRAS.422..926M}, the error in the linear method can easily reach thirty percent or more for the smallest scales. Thus one should expect significant errors in the quadrupole of the 21 cm power spectrum  computed with the linear approximation.

\begin{figure}
	\includegraphics[width=\columnwidth]{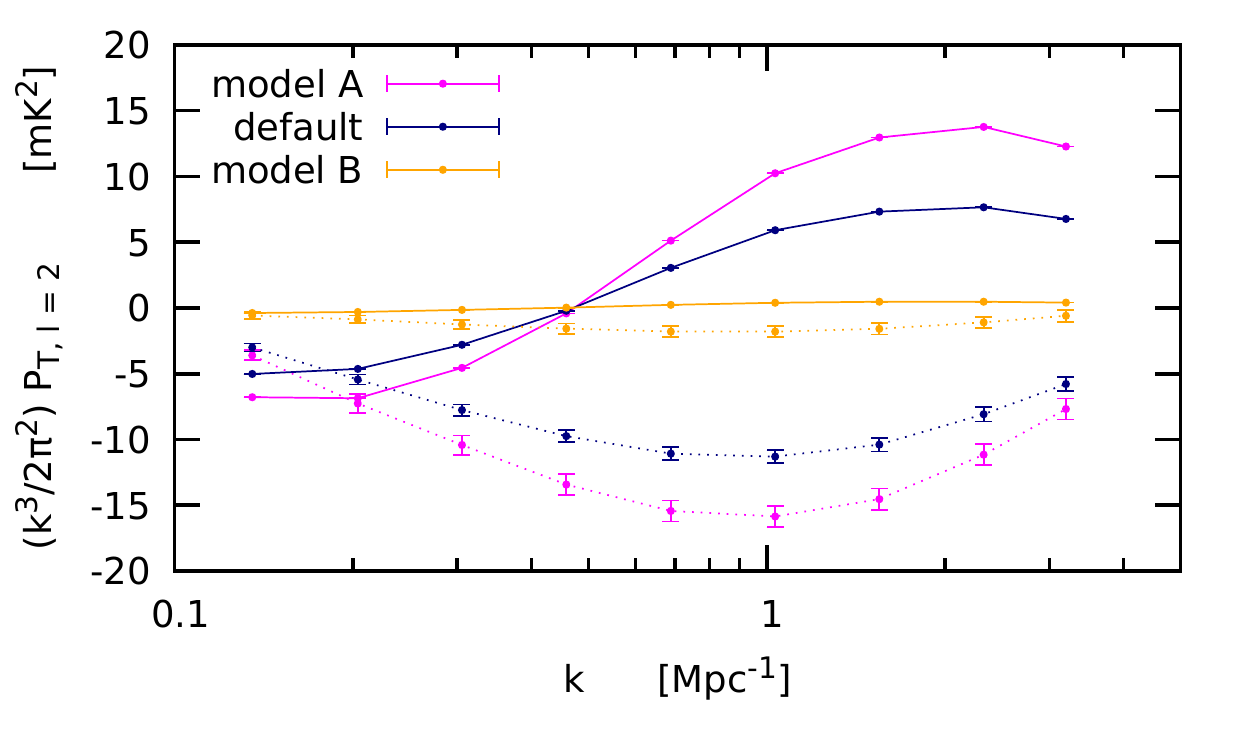}
    \caption{Failure of the linear $\mu_{\boldsymbol k}$-decomposition to reproduce the quadrupole of the 21 cm power spectrum at redshift 7 for the different models. The dashed lines represent the quadrupole from \texttt{21cmFAST} for the different models. In solid lines we have the quadrupole from the linear approximation.}
    \label{fig:QUADplot}
\end{figure}

We test the accuracy of the linear decomposition, i.e. we compare the right-hand side of Eq.~(\ref{eq:THEQUAD}) with the output from the \texttt{21cmFAST} simulations. We show the failure of the linear $\mu_{\boldsymbol{k}}$-decomposition in Fig.~\ref{fig:QUADplot} for the default model at redshift 7. The linear decomposition fails similarly for the other models.

Unfortunately, we see considerable errors in the decomposition. The main component of the discrepancy lies in the absence of non-linear effects (higher order correlations involving $\delta_{x_{\rm HI}}$ and $\delta_{m}$). This is not unexpected, as  \cite{2007ApJ...659..865L} and \cite{2012MNRAS.422..926M} had previously reported significant errors for the linear method due to absence of higher order correlations. In Fig.~\ref{fig:QUADplot} we see the effects of non-linear matter clustering and non-linear velocity perturbations at $z=7$, which cause the failure of the linear scheme. However, \cite{2012MNRAS.422..926M} also provided an alternative, the quasi-linear $\mu_{\boldsymbol k}$-decomposition. In the regime we are interested in, the quasi-linear method rewrites Eqs.~(\ref{eq:pmu0}--\ref{eq:pmu4}) into Eqs.(64--66) in \cite{2012MNRAS.422..926M}. In future work we will continue our efforts on extracting the quadrupole of the 21 cm power spectrum armed with the quasi-linear decomposition. 

Even though the $\mu_{\boldsymbol k}$-decomposition fails to reproduce the quadrupole of the 21 cm power spectrum, one can still see the correct trends at large scales. Specifically, the quadrupole from the earlier reionization model is larger than the one from the default model, and they are both larger than the quadrupole from the later reionization model. We take this trend as a positive sign that future work with the quasi-linear decomposition may be able to correctly mitigate the effect of patchy reionization in the \lya forest.

\section{Discussion}
\label{sec:summary}

Hydrogen reionization is one of the defining events in the thermal and dynamical history of the IGM. It heats up the IGM to $> 10^4 \ \textup{K}$, and by increasing the Jeans mass, it disrupts pre-existing small-scale structure. We have seen that the thermal and dynamical effects persist for cosmological timescales, and that the transmission of the Lyman-$\alpha$ forest even at $z_{\rm obs}<3$ is sensitive to the reionization model. Since reionization is believed to have occurred inhomogeneously -- by expanding and overlapping ionized ``bubbles'' correlated with large-scale structure -- this dependence on reionization redshift $z_{\rm re}$ translates into a spatial modulation of the Lyman-$\alpha$ forest and hence a correction to the power spectrum.

The magnitude of the effect is largest in the 3D power spectrum at the highest observed redshifts and on large scales (which are better matched to the scale of reionization bubbles). For example, in the 3D power spectrum at $z_{\rm obs}=4$ and $k=0.14$ Mpc$^{-1}$, we find corrections of 19--36\% depending on the reionization model chosen. At lower $z_{\rm obs}$, the effect of reionization is reduced, declining to 2.0--4.1\% at $z_{\rm obs}=2$. For the 1D \lya power spectrum the deviation is significantly smaller: again at $k=0.14$ Mpc$^{-1}$ we estimate 3.3--6.5\% at $z_{\rm obs}=4$, declining to 0.35--0.75\% at $z_{\rm obs}=2$. The corrections due to reionization are small, but we should remember that the 1D power spectrum is already measured at very high signal-to-noise ratio: for example at $z_{\rm obs}=2.2$ and $k=0.116$ Mpc$^{-1}$, BOSS+eBOSS have a statistical error of $1.2\%$ {\em per bin}\footnote{The bin size is $\Delta z=0.2$ and $\Delta k/k=0.03$, and corresponds to the second row of Table 4 in \citet{2018arXiv181203554C}.} \citep{2018arXiv181203554C}. These statistical errors will shrink further in the DESI era.

In principle, measurements of diffuse 21 cm radiation from the epoch of reionization can constrain the reionization model, one of the key inputs in calculating the correction to the \lya power spectrum. We are particularly interested in the 21 cm quadrupole $P_T^{\ell=2}(k,z)$, since it is sensitive to the specific power spectrum $P_{m,x_{\rm HI}}(k,z)$ that we need. Unfortunately, the simplest implementation of this idea -- the ``linear $\mu_{\boldsymbol k}$ decomposition'' theory -- is not accurate in the range of parameter space we need. In future work we will investigate other correction schemes, including models with corrections to linear theory (building on past work, e.g., \citealt{2012MNRAS.422..926M}), and schemes where reionization models are parameterized (e.g., source ionizing efficiency, minimum mass, IGM clumping parameters) and then 21 cm observations are used to constrain the parameters rather than directly infer power spectra.

Another avenue for future work is to incorporate some of the other physical effects that may interact with inhomogeneous hydrogen reionization. One of the most important may be He{\,\sc ii} reionization, which is believed to have occurred around $z\sim 3$ and resulted in an additional energy injection into the IGM. This has likely been seen in the thermal evolution of the IGM inferred from the \lya forest \citep[e.g.,][]{2011MNRAS.410.1096B, 2018arXiv180804367W}. This additional energy injection can reduce the sensitivity of the low-redshift IGM to its initial thermal state, e.g., it can reduce $\partial \ln T(z=z_{\rm obs})/\partial \ln T(z=z_{\rm re})$ (see, e.g., the discussion in \citealt{2018MNRAS.474.2173H} in the context of streaming velocities), though the change depends on the timeline and the relative contributions of EUV and X-ray radiation. Previous studies have also found that He{\,\sc ii} reionization introduces its own imprints on the \lya forest \citep{2009ApJ...694..842M, 2013MNRAS.435.3169C, 2015MNRAS.447.2503G}. In any case, it appears that at the level of precision of interest for modern cosmological \lya forest studies, the IGM may not have relaxed from inhomogeneous hydrogen reionization before helium reionization takes place.

A second issue is that we have taken only a simplified model for hydrogen reionization itself: we have ignored X-ray heating prior to hydrogen reionization, and we have neglected variations in the reheat temperature (e.g., due to spatial variation of the ionization parameter). The choice of modeling of these issues led to only minor changes in the simulations by \citet{2018MNRAS.474.2173H}, so we did not consider them in this paper, but only a few alternative models were tested and more should be explored. In addition, our small-scale simulations do not include fluctuations in the photoionization rate due to the clustering of the ionizing sources \citep{2014PhRvD..89h3010P,2014MNRAS.442..187G}, which will become especially relevant for the lower redshifts ($z<4$).

In conclusion, we have found that inhomogeneous hydrogen reionization results in an imprint on the \lya\ forest power spectrum, even at ``low'' redshifts $2<z_{\rm obs}<4$. The effect is present despite the ``attractor'' nature of the IGM temperature-density relation, because of the finite relaxation time and the low redshift of reionization favored by {\slshape Planck}. It can range from $\ll 1\%$ at small scales and low redshifts, up to tens of percents in the large-scale 3D power spectrum at $z_{\rm obs}\gtrsim 3.5$. While we have not yet developed a robust mitigation strategy, there are several clear paths forward on both the theory/simulation front, and with additional observations to help constrain reionization.

\section*{Acknowledgements}
We thank Hy Trac, Tzu-Ching Chang, Jordi Miralda-Escud\'e, Llu\'is Mas-Ribas, Jose O\~{n}orbe, Benjamin Buckman and Xiao Fang for useful discussions. PMC is grateful to Andrei Mesinger for fruitful suggestions with 21cmFAST. PMC and CMH are supported by the Simons Foundation, the US Department of Energy, the Packard Foundation, the NSF, and NASA. This material is based upon work supported by the U.S. Department of Energy, Office of Science, Office of High Energy Physics under Award Number DE-SC-0011726.




\bibliographystyle{mnras}
\bibliography{21cm} 



\appendix

\bsp	
\label{lastpage}
\end{document}